\documentclass[aps,showpacs,twocolumn,preprintnumbers,amsmath,amssymb]{revtex4}
\usepackage{graphicx}
\usepackage{bm}

\begin{document}

\title{Order parameter symmetry in ferromagnetic superconductors}
\author{K. V. Samokhin$^*$ and M. B. Walker}
\affiliation{Department of Physics, University
of Toronto, Toronto, Ontario, Canada M5S 1A7}
\date{\today}

\begin{abstract}
We analyze the symmetry and the nodal structure of the superconducting order
parameter in a cubic ferromagnet, such as ZrZn$_2$. We demonstrate how the
order parameter symmetry evolves when the electromagnetic interaction of the conduction 
electrons with the internal magnetic induction and the spin-orbit coupling are 
taken into account. These interactions break the cubic symmetry and  
lift the degeneracy of the order parameter. 
It is shown that the order parameter which appears immediately below the critical 
temperature has two components, and its symmetry is described by {\em co-representations} 
of the magnetic point groups. This allows us to make predictions about the location 
of the gap nodes.
\end{abstract}


\maketitle

\section{Introduction}
\label{sec:Intro}

A metallic ferromagnet is characterized by the fact that its
electronic energy bands are split by the exchange interaction
between the electrons so that the spin-up bands have different
energies from the spin-down bands. This has important consequences 
for the symmetry and the gap structure of possible 
superconducting states. 
In this article, we study the symmetry properties of the 
superconducting order parameter in 
a cubic ferromagnetic superconductor, such as ZrZn$_2$, in the limit of
weak spin-orbit coupling, thus complementing an earlier study by
the authors \cite{sam02} carried out in the strong spin-orbit
coupling limit.  

The formation of spin-singlet Cooper pairs in a
ferromagnet is strongly inhibited because electrons with opposite
momenta and spin have energies differing by the exchange splitting
of the energy bands \cite{cc_limit64}.  Therefore we consider here
only the case of spin-triplet pairing. In triplet
superconductivity, the order parameter has three components:
$\Delta_{\uparrow \uparrow}$ corresponding to the pairing of
electrons in the spin-up band, $\Delta_{\downarrow \downarrow}$
corresponding to the pairing of electrons in the spin-down
band, and $\Delta_{\uparrow \downarrow}$ corresponding to the
pairing of one spin-up and one spin-down electron.  The
$\Delta_{\uparrow \downarrow}$ component 
is expected to be very small for the same reasons that inhibit the
possibility of singlet superconductivity in a ferromagnet, and
thus we generally neglect it.  In the case of zero spin-orbit
coupling, there is no coupling between the three components of the
order parameter and thus, according to the Landau theory of second
order phase
transitions, only one of them will become nonzero immediately
below the superconducting transition temperature. The turning on
of a weak spin-orbit coupling has two effects: (i) there will be
changes to each of the three components of the order parameter
resulting from the lowering of the symmetry by the presence of the
spin-orbit interaction, and (ii) the three components of the order
parameter will be mixed together by the presence of the spin-orbit
interaction.  It will turn out that for the
ferromagnetic magnetization directed along any high-symmetry axis, 
and for all possible symmetries of
the superconducting gap function, at least one of the dominant 
components $\Delta_{\uparrow \uparrow}$ and 
$\Delta_{\downarrow\downarrow}$ has either line nodes or point nodes
in the momentum space. These zeros will become deep minima in the
energy gap in
the presence of the component $\Delta_{\uparrow \downarrow}$.  
The bulk of this paper is devoted to a detailed
demonstration of these results.

Recent discoveries of coexistence of superconductivity with itinerant
ferromagnetism in ZrZn$_2$ \cite{pfl01}, and UGe$_2$ \cite{sax00}
have renewed interest to the old problem of the interplay between the two
phenomena. These materials exhibit a number of peculiar properties.
First, in contrast to all previously known examples of ferromagnetic
superconductors, such as ternary rare-earth compounds, ruthenocuprates,
{\em etc}, the same band electrons ($d$-electrons in ZrZn$_2$, or
$f$-electrons in UGe$_2$) are responsible for both the superconductivity
and the ferromagnetism.
Second, the superconductivity occurs only in the ferromagnetic phase.
While the exchange splitting of the Fermi surfaces suppresses singlet
Cooper pairing, it was shown that the exchange by spin fluctuations
can lead to a triplet pairing both in the paramagnetic
and the ferromagnetic phases \cite{fay80}, or to the enhancement of the
superconducting critical temperature $T_c$ on the ferromagnetic side
\cite{kir01}. A prominent feature of the phase diagram of ZrZn$_2$ is 
that $T_c$ grows as pressure moves away from the ferromagnetic 
quantum critical point, which can be explained by the exchange-type 
interaction of the magnetic moments of the Cooper pairs with the 
magnetization density \cite{walker02}.

Even though the microscopic mechanism of pairing is not completely understood,
one can use symmetry analysis to identify the possible order
parameters and determine the structure of the superconducting gap.
The symmetry group ${\cal G}$ of the system in the normal state is defined as
a group of transformations which leave the system Hamiltonian $H_0$ invariant.
If the spin-orbit coupling is sufficiently strong, ${\cal G}$ contains the 
operations
which affect both the coordinate and the spin degrees of freedom.
In non-magnetic superconductors, time reversal
symmetry $K$ is not broken, and ${\cal G}=S\times K\times U(1)$,
where $S$ is the space group of the crystal, and $U(1)$ is the gauge group 
\cite{min99}. In magnetic superconductors, time reversal symmetry is broken,
and ${\cal G}=S_M\times U(1)$, where $S_M$ is the magnetic space group whose 
elements
leave both the microscopic charge density and the magnetization
density $\bm{M}$ invariant \cite{lan60}. For example, if there is a crystal
point group operation $R$ which transforms $\bm{M}$ to $-\bm{M}$,
then the combined operation $KR$ will be
an element of $S_M$, because time reversal restores the original
$\bm{M}$ not affecting the lattice symmetry.
The combined operation $KR$ is anti-linear and anti-unitary, which brings
about a number of novel features in the symmetry analysis compared to the
non-magnetic case.  The symmetry properties of
the superconducting state in ZrZn$_2$ assuming strong spin-orbit coupling
have been studied in Ref. \cite{sam02} (see also Refs. \cite{mac01},
\cite{fom01}, and \cite{min02}, where various aspects of the theory
of ferromagnetic superconductors have been considered).
However, a rather weak magnetic anisotropy in ZrZn$_2$
\cite{pfl01} points out that the spin-orbit coupling might be small,
which requires a modification of the analysis of Ref. \cite{sam02}. 
A peculiar feature of ferromagnetic superconductors, which was first
emphasized by Ginzburg \cite{ginz57}, is that the internal
magnetic induction in the normal state is always nonzero.
This means that the orbital motion of electrons and therefore the
symmetry of the 
superconducting order parameter will be affected by the ferromagnetic
magnetization even in the absence of spin-orbit coupling. Another
consequence is that the system undergoes the superconducting
phase transition into a mixed state, even in the absence of
an external field.

The paper is organized as follows. In Sec. \ref{sec:Deriv}, the
normal-state symmetry groups are derived assuming that spin-orbit coupling
is neglibly small, focusing on the
cubic crystal symmetry relevant for ZrZn$_2$.
In Sec. \ref{sec:SymmOP}, the effect of the electromagnetic interaction on
the symmetry of the spin-triplet
order parameters is analyzed, and the predictions are made about
the location of gap zeros. The lattice periodicity is taken into account
properly, which allows us to list all possible gap nodes, including
those at the surface of the first Brillouin zone.
In Sec. \ref{sec:SOEvolution}, the evolution of the
order parameter symmetry in the presence of spin-orbit coupling is studied, and
it is shown how the order parameter is induced on both sheets of the Fermi
surface. In Sec. \ref{sec:GL}, the Ginzburg-Landau free energy functionals
are derived for different magnetic symmetries.
Sec. \ref{sec:Concl} concludes with
a discussion of our results and their implications for the experiment.

\section{Derivation of the symmetry group at zero spin-orbit coupling}
\label{sec:Deriv}

We consider the case of cubic symmetry appropriate for ZrZn$_2$,
which has the cubic laves phase structure.  Also, we consider a
single spin-degenerate electron band which is split by an exchange
field in the ferromagnetic state. The symmetry of the normal (i.e.
non-superconducting) state will be analyzed in terms of the
effective single-particle Hamiltonian
\begin{eqnarray}
\label{Hamilt_0}
    H_0&=&\int d\bm{r}\;\psi^\dagger_\alpha(\bm{r})\biggl\{
    \frac{1}{2m}\left[-i\hbar\frac{\partial}{\partial\bm{r}}
    +\frac{e}{c}\bm{A}(\bm{r})\right]^2\delta_{\alpha\beta}
    \nonumber \\
    &&+U(\bm{r})\delta_{\alpha\beta}-[\bm{h}_{ex}(\bm{r})
    + g\mu_B\bm{B}]\cdot\bm{\sigma}_{\alpha\beta}\biggr\}\psi_\beta(\bm{r}).
\end{eqnarray}
Here $e$ is the absolute value of the electron charge,
$U(\bm{r})$ is the periodic crystal lattice potential, 
$\bm{\sigma}=(\sigma_1,\sigma_2,\sigma_3)$ are Pauli matrices, and
$\bm{h}_{ex}(\bm{r})$ is the exchange field.
The magnetic induction inside the ferromagnet (in the assumed long 
cylinder geometry) is $\bm{B}=\mathrm{curl}\,\bm{A}(\bm{r})=4\pi\bm{M}$, 
and $g$ is the Land\'{e} $g$-factor for electrons,
which determines the Zeeman splitting.  
In the case of a collinear ferromagnet, which is assumed here, 
$\bm{h}_{ex}(\bm{r})=\bm{h}_0 f(\bm{r})$, 
where $f(\bm{r})$ has the same periodicity as $U(\bm{r})$, and
$\bm{h}_0$ is the exchange field direction, which does not vary in
the crystallographic unit cell.
We also assume that $\bm{B}$ is uniform and there is no external magnetic
field (otherwise $\bm{B}=\bm{H}_{ext}+4\pi\bm{M}$), so that
the vector potential can, for example, be
written as $\bm{A}(\bm{r})=[\bm{B}\times\bm{r}]/2$. In
principle, the magnetic induction varies both in magnitude and direction 
in the crystallographic unit cell, and our $\bm{B}$ is the unit cell
average of the microscopic magnetic induction. If the variation of
the magnetic induction in the unit cell were taken into account it
would change the symmetry analysis given below. However, since
$\bm{B}$ (approximately 400 Gauss at zero pressure) is much
smaller than the exchange field in ZrZn$_2$, and since a spatial
average of magnetic induction over the unit cell is usually
assumed to be appropriate in the calculation of the effects of the
magnetic induction on the orbital motion of the electrons, the
approximation of a uniform $\bm{B}$ is sufficient.
The exchange field $\bm{h}_{ex}$, the magnetization density $\bm{M}$,
and the magnetic induction $\bm{B}$ all have a common
direction.  

The spin-orbit coupling in not included in Eq. (\ref{Hamilt_0}).
It should be noted that when we refer to spin-orbit coupling in
this article, we mean the single-particle spin-orbit coupling
which is shown explicitly in Eq. (\ref{H0so}) below. In principle,
the microscopic magnetic dipole-dipole interaction that gives rise to the
internal magnetic induction $\bm{B} = 4\pi\bm{M}$ couples the
spin and orbital motions, but because we assume a uniform $\bm{B}$
this does not affect our symmetry analysis.
Even in the absence of the spin-orbit coupling (\ref{H0so}), 
there is an effect of the ferromagnetic magnetization 
density on the orbital motion of the electrons, which we refer to as the
electromagnetic interaction. This means that 
the symmetry and the free energy of the superconducting
state will depend on the direction of $\bm{M}$.  

At zero spin-orbit coupling, the symmetry operations act independently
in the real (orbital) space and the spin space, so that
the full symmetry group of $H_0$ is a direct product
\begin{equation}
\label{Gfull}
  {\cal G}=\mathbf{G}_{orb}\times\mathbf{G}_{spin}\times U(1),
\end{equation}
where $U(1)$ is the gauge group composed of phase rotations
$\Phi\psi_\alpha(\bm{r})\Phi^{-1}=e^{i\phi}\psi_\alpha(\bm{r})$.
For our purposes, the translations are
not important, so that $\mathbf{G}_{orb}$ contains orbital rotations $R_{orb}$:
\begin{equation}
\label{orb_rots}
  R_{orb}\psi_\alpha(\bm{r})R_{orb}^{-1}=
   \psi_\alpha(R_{orb}^{-1}\bm{r}),
\end{equation}
and inversion:
\begin{equation}
\label{Idef}
     I\psi_\alpha(\bm{r})I^{-1}=\psi_\alpha(-\bm{r}).
\end{equation}
Also, the effects of time-reversal symmetry are included in
$\mathbf{G}_{orb}$.  Below we shall use the notation $C_{k\bm{n}}$
for the rotations by an angle $2\pi/k$ about the axis $\hat{\bm{n}}$ in
orbital space. The group $\mathbf{G}_{spin}$ contains spin
rotations $R_{spin}$:
\begin{equation}
\label{spin_rots}
  R_{spin}\psi_\alpha(\bm{r})R_{spin}^{-1}=
  [D^{(1/2)}(R_{spin})]_{\alpha\beta}\psi_{\beta}(\bm{r}),
\end{equation}
where $D^{(1/2)}(R)$ is the spinor ($j=1/2$) representation of rotations:
for a rotation $R$ by an angle $\theta$ around $\hat{\bm{n}}$,
$D^{(1/2)}(R)=\exp[-i(\theta/2)(\bm{\sigma}\cdot\hat{\bm{n}})]$.
It is convenient to introduce an orthogonal basis of unit vectors
$\hat{\bm{e}}_1,\hat{\bm{e}}_2,\hat{\bm{e}}_3$ in the spin space, such that
$\hat{\bm{e}}_3\parallel\bm{B}$. We shall use the notation $C^{s}_{k\bm{n}}$
for the rotations by an angle $2\pi/k$ about the axis $\hat{\bm{n}}$ 
in spin space.

A standard representation for the time reversal operator $K$ is
$K=C^s_{2e_2}K_0$, where $K_0$ is the complex conjugation operator
associated with the representation $\{\bm{r}, s_z\}$
\cite{messiah}. The anti-unitary operator $K_0$ is defined more
explicitly by the equation
\begin{equation}
    K_0 [c\psi_\alpha(\bm{r})] K_0 = c^*\psi_\alpha(\bm{r}),
\label{K_0}
\end{equation}
where $c$ is an arbitrary $c$-number. In the momentum representation,
$K_0$ also reverses the sign of $\bm{k}$.
It should be noted that, in the decomposition $K=C^s_{2e_2}K_0$,
while $C^s_{2e_2}$ is an operator in spin space only, $K_0$
operates in both spin and orbital space [as indicated, for
example, by the result $K_0s_yK_0 = -s_y$, where $s_y=(\hbar/2)\sigma_2$
is the $y$-component of the electron spin operator].  
Nevertheless, in discussing the symmetry properties of 
the Hamiltonian $H_0$ given by Eq. (\ref{Hamilt_0})
for the case where the common direction of
$\bm{h}_{ex}$ and $\bm{B}$ is along $\hat{\bm{e}}_3$ (so that the
Hamiltonian does not contain $\sigma_2$), it is useful to consider
$K_0$ together with the symmetry operations in orbital space.

If $R_{orb}$ leaves the periodic potential $U(\bm{r})$ and the
exchange field $\bm{h}_{ex}(\bm{r})$ 
invariant, then the transform of the  Hamiltonian $H_0$, namely
$R_{orb} H_0 R_{orb}^{-1}$, is the same as $H_0$ except that the
vector potential $\bm{A}(\bm{r})$ is replaced by
$\bm{A}'(\bm{r})=R_{orb}^{-1}\bm{A}(R_{orb}\bm{r})$. This means
that the transformation rule for the magnetic induction under 
$R_{orb}$ is $\bm{B}(\bm{r})=\mathrm{curl}\,\bm{A}(\bm{r})
\to\bm{B}'(\bm{r})=\mathrm{curl}\,\bm{A}'(\bm{r})=
R_{orb}^{-1}\bm{B}(R_{orb}\bm{r})=R_{orb}^{-1}\bm{B}(\bm{r})$. 
Also, $K_0 H_0 K_0$ is the
same as $H_0$ except that $\bm{A}(\bm{r})$ is replaced by 
$-\bm{A}(\bm{r})$, so that the transformation rule for the 
magnetic induction under $K_0$ is simply $\bm{B}\to-\bm{B}$.  
Thus, if $R_{orb}$ leaves $U(\bm{r})$ and $\bm{h}_{ex}(\bm{r})$
invariant, and $R_{orb}\bm{B}=-\bm{B}$, then $K_0R_{orb}$ is a
member of the symmetry group of $H_0$. For convenience, such
combined symmetry elements will be included with purely orbital
elements in the definitions of the various orbital symmetry groups
below.

In the non-magnetic case, i.e. at $\bm{h}_{ex} = \bm{B} =
\bm{M}=0$, the orbital symmetry of $H_0$ is determined by the
symmetry of the lattice potential $U(\bm{r})$. Since ZrZn$_2$ has
a cubic Laves phase structure, $\mathbf{G}_{orb}=\mathbf{O}_h
\times \mathbf{K}_0=\mathbf{O}\times\mathbf{I}\times\mathbf{K}_0$, 
where $\mathbf{I} = \{E, I \}$ and $\mathbf{K}_0 = \{E, K_0\}$. In
addition, $H_0$ is invariant under arbitrary rotations in the spin
space, so that $\mathbf{G}_{spin}=SU(2)$.

In the ferromagnetic case, where $\bm{M}$, $\bm{h}_{ex}$ and
$\bm{B}$ are all nonzero,  time reversal symmetry is broken, and,
as noted above, the symmetry group of $H_0$ contains  elements of
the form $K_0R_{orb}$ as well as purely orbital transformations.
In addition, the symmetry group of $H_0$ will contain operations
that are purely spin-space rotations.  More precisely, it is
evident from Eqs. (\ref{Hamilt_0}) and (\ref{spin_rots}) that
$H_0$ is invariant under the operators of the group
$\mathbf{C}_{\infty e_3}^s$ describing the set of all spin
rotations about the axis $\hat{\bm{e}}_3$, which, as always, is taken to
lie along the common direction of $\bm{M}$, $\bm{h}_{ex}$ and
$\bm{B}$. Therefore,
\begin{equation}
\label{Gspin}
    \mathbf{G}_{spin}= \mathbf{C}_{\infty e_3}^s.
\end{equation}
This spin-space symmetry group will be combined with a number of
orbital symmetry groups to describe a number of different cases
corresponding to different orientations for the ferromagnetic
magnetization density.  The different cases will be called Case A,
Case B, ... Case E. The appropriate symmetry groups will be
described immediately, and the order-parameter symmetries for each
of the cases will be described later in section\ \ref{sec:SymmOP}.

\paragraph*{Case A.} The orbital symmetry of the system is determined by
the electromagnetic interaction of the conduction electrons with
the induction $\bm{B}$ via the vector potential $\bm{A}$. If this
interaction can be neglected, which amounts to setting $e\to 0$ in
Eq. (\ref{Hamilt_0}), then the Hamiltonian is real, so that the
orbital symmetry is independent of $\bm{M}$ and is described by
the cubic group $\mathbf{O}_h$, i.e.
\begin{equation}
\label{GorbCaseA}
  \mathbf{G}_{orb}=\mathbf{O}_h \times \mathbf{K}_0
            =\mathbf{O}\times \mathbf{I} \times \mathbf{K}_0.
\end{equation}
In this case, which might be appropriate for a neutral Fermi system, 
such as the liquid $^3$He in magnetic field, or the ``cold'' atomic gases,
only the spin symmetry is influenced by the presence of ferromagnetic
magnetization.

In ferromagnet metals, the electromagnetic interaction is always present, and
the presence of the magnetization affects the orbital symmetry
even in the absence of spin-orbit coupling.  The structure of the
orbital group depends on the direction of
magnetization density. In ZrZn$_2$ the magnetic anisotropy is
sufficient weak that it should be possible to align the
magnetization density along an arbitrary direction in the crystal
by applying an external magnetic field along that direction.  We now
consider a number of possible orientations.

\paragraph*{Case B.}  If the magnetization density lies along the
[001] direction, the orbital symmetry group is
\begin{eqnarray}
\label{GorbCaseB}
  \mathbf{G}_{orb}&=&\mathbf{D}_4(\mathbf{C}_4)\times \mathbf{I}
    =\{E,C_{4z},C_{2z},C^{-1}_{4z},
  \nonumber \\
  && K_0C_{2x},K_0C_{2y},K_0C_{2a},K_0C_{2b}\}\times \mathbf{I},
\end{eqnarray}
where $\hat{\bm{a}}=(\hat{\bm{x}}+\hat{\bm{y}})/\sqrt{2}$, and
$\hat{\bm{b}}=(\hat{\bm{x}}-\hat{\bm{y}})/\sqrt{2}$. Here we use a
standard notation for the magnetic group $\mathbf{G}(\mathbf{H})$
\cite{magn_groups}, where the subgroup $\mathbf{H}$ in parentheses
(the unitary subgroup) includes all  elements of $\mathbf{G}$
which are not multiplied by the anti-unitary operation $K_0$.
A useful observation is that any magnetic group
$\mathbf{G}(\mathbf{H})$ can be expressed in terms of left
cosets with respect to the unitary subgroup $\mathbf{H}$:
$\mathbf{G}(\mathbf{H})=\mathbf{H}+A\mathbf{H}$, where all
elements of the coset $A\mathbf{H}$ are anti-unitary. The choice 
of the anti-unitary group element $A$ is arbitrary and does not 
affect the final results, but once chosen it remains fixed.
For the group $\mathbf{D}_4(\mathbf{C}_4)$, we choose $A=K_0C_{2x}$.

\paragraph*{Case C.}  When the magnetization density lies along
the [111] direction, the orbital symmetry group is
\begin{eqnarray}
\label{GorbCaseC}
  \mathbf{G}_{orb}&=&\mathbf{D}_3(\mathbf{C}_3)\times \mathbf{I}=
  \{E,C_{3\epsilon},C^{-1}_{3\epsilon},\nonumber \\
  && K_0C_{2b},K_0C_{2b'},K_0C_{2b''}\}\times \mathbf{I},
\end{eqnarray}
where
$\hat{\bm{\epsilon}}=(\hat{\bm{x}}+\hat{\bm{y}}+\hat{\bm{z}})/\sqrt{3}$,
$\hat{\bm{b}}'=C_{3\epsilon}\hat{\bm{b}}=(\hat{\bm{y}}-\hat{\bm{z}})/\sqrt{2}$,
and
$\hat{\bm{b}}''=C^{-1}_{3\epsilon}\hat{\bm{b}}=
(\hat{\bm{z}}-\hat{\bm{x}})/\sqrt{2}$. For this magnetic group, we
choose $A=KC_{2b}$.

\paragraph*{Case D.}
When the magnetization density lies along the [110] direction, the
orbital symmetry group is
\begin{equation}
\label{GorbCaseD}
  \mathbf{G}_{orb}=\mathbf{D}_2(\mathbf{C}_2)\times \mathbf{I}
    =\{E,C_{2a},K_0C_{2b},K_0C_{2z}\}\times \mathbf{I}.
\end{equation}
In this case, we also choose $A=KC_{2b}$.

\paragraph*{Case E.} For the magnetization along a general
direction, the orbital symmetry group is
\begin{equation}
\label{GorbCaseE}
    \mathbf{G}_{orb}=\mathbf{C}_i=\mathbf{C}_1\times\mathbf{I},
\end{equation}
where $\mathbf{C}_1$ consists of the unity operation $E$.
This group is trivial and does not contain anti-unitary elements.

In the next section, we study the symmetry properties of the
superconducting order parameter at $\bm{M}\neq 0$ using
$\mathbf{G}_{spin}$ from Eq. (\ref{Gspin}), and $\mathbf{G}_{orb}$
from Eqs.
(\ref{GorbCaseA},\ref{GorbCaseB},\ref{GorbCaseC},\ref{GorbCaseD},
\ref{GorbCaseE}). The microscopic origins of the ferromagnetism
and the superconductivity are not important for the symmetry
analysis.

\section{Superconducting order parameter at zero spin-orbit coupling}
\label{sec:SymmOP}

In ZrZn$_2$, the exchange band splitting is $E_{ex}\simeq
5\mathrm{mRy}\simeq 800\mathrm{K}$ \cite{santi01}, which greatly
exceeds the superconducting critical temperature $T_c\simeq
0.2\mathrm{K}$. In this conditions, the usual
Chandrasekhar-Clogston arguments \cite{cc_limit64} make any
pairing of electrons with opposite spins, in particular in the
singlet channel, strongly suppressed. The general form of a
spin-triplet superconducting order parameter is
$\Delta_{\alpha\beta}(\bm{k},\bm{r})=(i\bm{\sigma}\sigma_2)_{\alpha\beta}
\bm{d}(\bm{k},\bm{r})$ \cite{min99}. It is convenient to
use the following representation:
$\bm{d}(\bm{k})=\hat{\bm{e}}_+d_-(\bm{k})+\hat{\bm{e}}_-d_+(\bm{k})+
\hat{\bm{e}}_3d_3(\bm{k})$, where $\hat{\bm{e}}_\pm=(\hat{\bm{e}}_1\pm
i\hat{\bm{e}}_2)/\sqrt{2}$ and $d_\pm=(d_1\pm id_2)/\sqrt{2}$.

According to the Landau
theory of phase transitions, the spin vector 
$\bm{d}$, which appears at the critical temperature $T_c$, should
correspond to an irreducible representation of the normal state
symmetry group $\cal{G}$. The easiest way to obtain the 
transformation properties of the order parameter under the operations 
from $\cal{G}$, i.e. the orbital and the spin rotations, and also the operation 
$K_0$, is to use the mean-field expression for the pairing Hamiltonian: 
\begin{equation}
\label{H_sc}
     H_{sc}=\sum\limits_{\bm{k}}\sum\limits_{\alpha,\beta=\uparrow,\downarrow}
     \Bigl[\Delta_{\alpha\beta}(\bm{k})c^\dagger_{\bm{k}\alpha}
     c^\dagger_{-\bm{k},\beta}+\mathrm{H.c.}\Bigr].
\end{equation}
Here $\Delta_{\uparrow\uparrow}=-\sqrt{2}\,d_-$, which corresponds to a gap on the spin-up
Fermi surface; $\Delta_{\downarrow\downarrow}=\sqrt{2}\,d_+$, which corresponds to a
gap on the spin-down Fermi surface; and $\Delta_{\uparrow\downarrow}=
\Delta_{\downarrow\uparrow}=d_3$, 
which corresponds to a pairing of a spin-up electron with a spin-down electron.
Because of the Pauli principle, $\bm{d}(-\bm{k})=-\bm{d}(\bm{k})$.
From Eqs. (\ref{H_sc}), (\ref{orb_rots}), (\ref{spin_rots}), and (\ref{K_0}),
we obtain
\begin{eqnarray}
\label{d_rules}
 &&R_{orb} d_\alpha(\bm{k})=d_\alpha(R_{orb}^{-1}\bm{k})\nonumber\\
 &&R_{spin} d_\alpha(\bm{k})=[D^{(1)}(R_{spin})]_{\alpha\beta}d_\beta(\bm{k})\\
 &&K_0 d_\alpha(\bm{k})=d^*_\alpha(-\bm{k})\nonumber,
\end{eqnarray}
where $\alpha=\pm,3$, and $D^{(1)}(R)$ is the vector ($j=1$) representation of 
rotations. 

Since $\cal{G}$ is
a direct product of the independent orbital and spin symmetry groups 
(\ref{Gfull}),
the basis functions of the irreducible representations of $\cal{G}$ are 
given by products of the basis functions of $\mathbf{G}_{orb}$ and 
$\mathbf{G}_{spin}$.
An important point here is that, because of the presence of the 
anti-unitary operations $K_0R_{orb}$ in $\mathbf{G}_{orb}$, the symmetry 
analysis should be modified. The order parameter should transform according
to one of the irreducible {\em co-representations} of 
$\mathbf{G}_{orb}=\mathbf{G}(\mathbf{H})$,
which can be derived from the irreducible representations of the unitary 
subgroup $\mathbf{H}$ \cite{magn_groups}.

At $\bm{M}=0$, $\mathbf{G}_{spin}=SU(2)$, and $\bm{d}$ transforms
according to the three-dimensional vector representation of
$SU(2)$, whose basis functions are $\hat{\bm{e}}_\pm$ and $\hat{\bm{e}}_3$.
All three spin components $d_\pm$ and $d_3$ have the same critical 
temperature. 
At $\bm{M}\neq 0$, the spin
symmetry is reduced to $\mathbf{G}_{spin}= \bm{C}^s_{\infty e_3}$
[see Eq. (\ref{Gspin})], and the vector representation is split
into three one-dimensional representations of the group 
$\bm{C}^s_{\infty e_3}$. The spin components $d_\pm$ and $d_3$ have different
critical temperatures, and we assume that the maximum $T_c$ is
achieved for $d_-$. Thus, the order parameter can be represented
as an expansion
\begin{equation}
\label{dreduced}
  \bm{d}_\Gamma(\bm{k},\bm{r})=
  i\hat{\bm{e}}_+ \sum\limits_{i=1}^{n_\Gamma}\eta_i(\bm{r})
   f_{\Gamma,i}(\bm{k}).
\end{equation}
Here $f_{\Gamma,i}(\bm{k})$ are the odd basis functions of a
$n_\Gamma$-dimensional irreducible co-representation $\Gamma$ of
$\mathbf{G}_{orb}$ (the parity of the spin-triplet order parameter
is fixed, and the inversion operation can be omitted from
$\mathbf{G}_{orb}$). The action of the orbital symmetry elements
on the functions $f_{\Gamma,i}(\bm{k})$ in the momentum space is
defined as follows: under the crystal rotations,
$R_{orb}f(\bm{k})=f(R_{orb}^{-1}\bm{k})$, under the combined
operations, $K_0R_{orb}f(\bm{k})=f^*(-R_{orb}^{-1}\bm{k})$. The
expansion coefficients $\eta_i(\bm{r})$ play the role of the order
parameter components, which enter the Ginzburg-Landau free energy
functional. The  factor $i$ on the right-hand side of Eq.
(\ref{dreduced}) is introduced so that, as we shall see in Sec.
\ref{sec:SOEvolution}, the anti-unitary combined operations $KR$
are equivalent to complex conjugation when acting on $\eta_i$.

The physical meaning of Eq. (\ref{dreduced}) is that the order parameter 
appears only on the spin-up sheet of the Fermi surface, while the 
spin-down sheet remains normal (for the order parameter on the 
spin-down sheet, one would have $d_+\neq 0$, i.e.
$\bm{d}\propto\hat{\bm{e}}_-$). 
It should be mentioned here that the band structure of
ZrZn$_2$ is quite complex \cite{maz97,santi01}, but we neglect such
complication here and assume that there are only two exchange-split bands. 
This assumption should not affect the essence of our results.
In contrast to the strong spin-orbit coupling case considered in Ref. 
\cite{sam02}, the interband interactions 
$c^\dagger_{\bm{k}\uparrow}c^\dagger_{-\bm{k},\uparrow}
c_{\bm{k}'\downarrow}c_{-\bm{k}',\downarrow}$, which could induce the order
parameters of the same symmetry on both sheets of the Fermi surface, are 
absent due to the spin conservation.
The critical temperature for the order parameter $d_3$, which describes the 
Cooper pairing of electrons with opposite spins,
is expected to be much smaller than those for $d_\pm$, because of the large 
value of the exchange splitting in ZrZn$_2$, mentioned in the beginning 
of this Section. For the same reason, we also neglect the
possibility of a superconducting state with a non-zero momentum, i.e. with
$\langle c^\dagger_{\bm{k}+\bm{q},\uparrow}
c^\dagger_{-\bm{k},\downarrow}\rangle\neq 0$
(Larkin-Ovchinnikov-Fulde-Ferrell state) \cite{loff}.

When the symmetry is described by one of the magnetic point groups
(\ref{GorbCaseB}), (\ref{GorbCaseC}), (\ref{GorbCaseD}), or
(\ref{GorbCaseE}),
$\mathbf{G}_{orb}$ has only one-dimensional co-representations
(see below), therefore Eq. (\ref{dreduced}) reduces to the form
\begin{equation}
\label{d1D}
  \bm{d}_\Gamma(\bm{k},\bm{r})=i\hat{\bm{e}}_+ \eta(\bm{r})f_\Gamma(\bm{k}).
\end{equation}
Thus, the order parameter has one component, and the Ginzburg-Landau functional
has the same form as for the conventional $s$-wave pairing. This means that the 
phase
transition from the normal ferromagnetic state to the superconducting state 
occurs
into the usual mixed state with a lattice of the Abrikosov vortices.
However, in contrast to the $s$-wave case, the orbital symmetry is
non-trivial, in particular, there are zeros in the spectrum
of elementary excitations where $f_\Gamma(\bm{k})=0$.
Below we examine the order parameter symmetry for different cases
and determine the positions of the gap zeros
dictated by the magnetic symmetry.

\subsection*{Case A:\quad $\mathbf{G}_{orb}=\mathbf{O}\times \mathbf{I} \times 
\mathbf{K}_0$}
\label{sec:CaseA}

\begin{table}
\caption{\label{table1} The examples
of the odd basis functions for the irreducible
representations of the point group
$\mathbf{O}$ from Ref. \cite{vol85}, $\omega=e^{2\pi i/3}$.}
\begin{ruledtabular}
\begin{tabular}{|c||c|}
   $\Gamma$  & $f_\Gamma(\bm{k})$ \\ \hline
   $A_1$      &  $k_xk_yk_z(k_x^2-k_y^2)(k_y^2-k_z^2)(k_z^2-k_x^2)$\\ \hline
   $A_2$      &  $k_xk_yk_z$\\ \hline
   $E$  &  $k_xk_yk_z(k_x^2+\omega k_y^2+\omega^*k_z^2),
           k_xk_yk_z(k_x^2+\omega^*k_y^2+\omega k_z^2)$ \\ \hline
   $F_1$  &  $k_x,k_y,k_z$ \\ \hline
   $F_2$  &  $k_x(k_y^2-k_z^2),k_y(k_z^2-k_x^2),k_z(k_x^2-k_y^2)$
\end{tabular}
\end{ruledtabular}
\end{table}

In this case, which is relevant for the superconductivity in a neutral 
ferromagnetic
Fermi system, the orbital symmetry is not affected by the presence of
a non-zero $\bm{M}$.
The order parameter is given by Eq.~(\ref{dreduced}).
The group $\mathbf{O}$ has 2 one-dimensional ($A_1$ and $A_2$), 1 two-
dimensional ($E$),
and 2 three-dimensional ($F_1$ and $F_2$) representations.
The examples of the basis functions are given in Table \ref{table1}.
One-component order parameters $\bm{d}_{A_1}(\bm{k})$ and
$\bm{d}_{A_2}(\bm{k})$ have line zeros at the Fermi surface, which
do not depend on the choice of the basis functions.
For the higher-dimensional representations, the form of the order parameter
and its gap structure are obtained by minimizing the free energy
in the superconducting state. The explicit expressions for the Ginzburg-Landau
functionals and the phase diagrams for the multi-component order parameters
can be found, e.g. in Ref. \cite{min99}.

In a charged Fermi system, where the vector potential created by the internal
magnetization affects the single-electron wave functions,
the cubic symmetry is reduced to one of the magnetic
groups (\ref{GorbCaseB}), (\ref{GorbCaseC}),
or (\ref{GorbCaseD}), and the degeneracy of
the two- and three-dimensional order parameters is lifted. Mathematically, this
corresponds to the splitting of higher-dimensional representations of 
$\mathbf{O}$
into several one-dimensional co-representations. If
$\bm{M}\parallel[001]$ and $\mathbf{O}\to\mathbf{D}_4(\mathbf{C}_4)$,
then it is easy to check, using Table \ref{table2}, that
\begin{eqnarray}
 && A_1 \to A \nonumber\\
 && A_2 \to B \nonumber\\
 && E \to A+B\\
 && F_1 \to A+{}^1E+{}^2E \nonumber\\
 && F_2 \to B+{}^1E+{}^2E \nonumber.
\end{eqnarray}
We also gave here the correspondence between the one-dimensional
representations of $\mathbf{O}$ and the co-representations of
$\mathbf{D}_4(\mathbf{C}_4)$. If
$\bm{M}\parallel[111]$ and $\mathbf{O}\to\mathbf{D}_3(\mathbf{C}_3)$, then,
using Table \ref{table3},
\begin{eqnarray}
 && A_1 \to A \nonumber\\
 && A_2 \to A \nonumber\\
 && E \to {}^1E+{}^2E\\
 && F_1 \to A+{}^1E+{}^2E \nonumber\\
 && F_2 \to A+{}^1E+{}^2E \nonumber.
\end{eqnarray}
If $\bm{M}\parallel[110]$ and $\mathbf{O}\to\mathbf{D}_2(\mathbf{C}_2)$, then,
using Table \ref{table4},
\begin{eqnarray}
 && A_1 \to A \nonumber\\
 && A_2 \to B \nonumber\\
 && E \to A+B\\
 && F_1 \to A+B+B \nonumber\\
 && F_2 \to A+A+B \nonumber.
\end{eqnarray}

The physical origin of the order parameter splitting
can be easily traced using the phenomenological Ginzburg-Landau theory.
For example, consider an uncharged Fermi liquid as above and let
$\bm{\eta}=(\eta_x,\eta_y,\eta_z)$ be a three-component order parameter
corresponding to the vector representation $F_1$ of the orbital group
$\mathbf{G}_{orb} = \mathbf{O} \times \mathbf{I} \times \mathbf{K}_0$ and
corresponding to a gap function on the spin-up Fermi surface.
Then the Ginzburg-Landau free energy describing a homogeous phase is
\begin{eqnarray}
    F &= & \alpha \bm{\eta}^\ast \cdot \bm{\eta}
        + \beta_1 (\bm{\eta}^\ast \cdot \bm{\eta})^2\nonumber \\
      && + \beta_2 |\bm{\eta} \cdot \bm{\eta}|^2
        + \beta_3 (|\eta_x|^4 + |\eta_y|^4 + |\eta_z|^4),
\label{GLuncharged}
\end{eqnarray}
where $\alpha=a(T-T_{c,0})$, and $T_{c,0}$ is the critical
temperature at $e=0$.
There are a number of physically different states that minimize this
free energy, depending on the values of the parameters
of the fourth-order terms \cite{vol85}; for example, one
of these solutions has the form
$\bm{\eta}=\eta_0 (1,1,1)$.

Now, for a charged (metallic)
ferromagnet, it is important to include the gradient terms in the free energy, 
so that the terms in the free energy of second order in the order parameter
becomes \cite{sigr91}
\begin{eqnarray}
  F &=& a(T-T_{c,0})|\bm{\eta}|^2+K_1(D_i\eta_j)^*(D_i\eta_j)\nonumber \\
   && +K_2(D_i\eta_i)^*(D_j\eta_j)+K_3(D_i\eta_j)^*(D_j\eta_i)\nonumber \\
   &&+K_4(D_i\eta_i)^*(D_i\eta_i)\nonumber\\
   &=& a(T-T_{c,0})|\bm{\eta}|^2+i\gamma [\bm{\eta}^*\times\bm{\eta}]\bm{B}+
  K_1(D_i\eta_j)^*(D_i\eta_j)\nonumber \\
  && +K_{23}[(D_i\eta_i)^*(D_j\eta_j)+(D_i\eta_j)^*(D_j\eta_i)]
  \nonumber\\
  && +K_4(D_i\eta_i)^*(D_i\eta_i).
\label{GLvector}
\end{eqnarray}
Here $\bm{D}=\bm{\nabla}+i(2\pi/\Phi_0)\bm{A}$, $\Phi_0=\pi\hbar c/e$ is the
flux quantum, $K_{23}=(K_2+K_3)/2$, and $\gamma=\pi(K_3-K_2)/\Phi_0$.
In the second part of Eq. (\ref{GLvector}), we regrouped the gradient terms
using the identity $[D_i,D_j]=-(2\pi i/\Phi_0)e_{ijk}B_k$.
The quantity $i[\bm{\eta}^*\times\bm{\eta}]$ can be interpreted, up to a factor,
as the density of the orbital magnetization of Cooper pairs \cite{min99}.
The second order terms given by Eq.\ (\ref{GLvector}) are sufficient to
calculate the critical temperature describing the phase transition from
the normal state to the superconducting mixed state.
The free energy of the superconducting state will
depend on the direction of the flux lines (determined by the direction
of $\bm{M}$ relative to the underlying crystal lattice).

Here we consider only the case $\bm{M}\parallel[001]$, so that
$\mathbf{G}_{orb}=\mathbf{D}_4(\mathbf{C}_4)$.
The critical temperature for the order parameter component $\eta_z$ can be
calculated exactly, while the critical temperatures for $\eta_\pm=\eta_x\pm 
i\eta_y$
can be found using the variational approach similar to that of
Ref. \cite{machida85}, with the result:
\begin{eqnarray}
\label{Tcs}
  &&T_{c,+}=T_{c,0}-
\frac{8\pi^2}{a\Phi_0}\left(K_1+K_3+\frac{K_4}{2}\right)M\nonumber\\
  &&T_{c,-}=T_{c,0}-\frac{8\pi^2}{a\Phi_0}\left(K_1+K_2+\frac{K_4}{2}\right)M\\
  &&T_{c,z}=T_{c,0}-\frac{8\pi^2}{a\Phi_0}K_1M. \nonumber
\end{eqnarray}
Barring accidental degeneracies, these critical temperatures are all different, 
so that,
at $e\neq 0$, the three-component order parameter is split.
The difference between the critical temperatures $T_{c,+}$ and $T_{c,-}$
is proportional to $\gamma$, and is entirely due to the interaction of the 
orbital pair
magnetization with $\bm{B}$.
It is easy to see, using Table \ref{table2}, that the order parameter
components $\eta_+$, $\eta_-$, and $\eta_z$ correspond to the following one-
dimensional
co-representations of $\mathbf{D}_4(\mathbf{C}_4)$: $\eta_z\sim A$, 
$\eta_+\sim{}^1E$,
$\eta_-\sim{}^2E$. It may be that as the temperature is lowered below this
critical temperature into the superconducting state and the fourth-order terms 
in the free energy become more important, there will be a second phase transition 
that into a state that does a better job of minimizing the fourth-order 
contributions to the free energy.

\subsection*{Case B:\quad $\mathbf{G}_{orb}=\mathbf{D}_4(\mathbf{C}_4)
             \times\mathbf{I}$\quad ($\bm{M}\parallel[001]$)}
\label{sec:CaseB}

\begin{table}
\caption{\label{table2} The character table and the examples
of the odd basis functions for the irreducible
co-representations of the magnetic point group
$\mathbf{D}_4(\mathbf{C}_4)$.
The overall phases of the basis functions are chosen so that
$K_0C_{2x}f_\Gamma(\bm{k})=f_\Gamma(\bm{k})$.
$\lambda_{1,2}$ are arbitrary real constants.}
\begin{ruledtabular}
\begin{tabular}{|c||c|c|c|}
   $\Gamma$  & $E$ & $C_{4z}$ &  $f_\Gamma(\bm{k})$ \\ \hline
   $A$      & 1   & 1  &  $k_z$  \\ \hline
   $B$      & 1   & $-1$ &  $k_z[\lambda_1(k_y+ik_x)^2+
            \lambda_2(k_y-ik_x)^2]$\\\hline
   $^1E$  & 1   & $i$ &  $k_y+ik_x$ \\ \hline
   $^2E$  & 1   & $-i$ &  $k_y-ik_x$
\end{tabular}
\end{ruledtabular}
\end{table}

The order parameter is given by Eq. (\ref{d1D}), and the
irreducible co-representations are listed in Table \ref{table2}.
We see that the order parameters $\bm{d}_A(\bm{k})$ and
$\bm{d}_B(\bm{k})$ vanish on the line $k_z=0$ at the Fermi
surface, while $\bm{d}_{^1E}(\bm{k})$ and $\bm{d}_{^2E}(\bm{k})$
vanish at the points $k_z=k_y=0$ [note that here the label
$\Gamma$ refers to the orbital symmetry, whereas in Ref.
\cite{sam02} we labelled the co-representations by their total
(orbital plus spin) symmetry]. These zeros are not accidental in
the sense that they are independent of the choice of the basis
functions. Indeed, one of the elements of the magnetic point group
$\mathbf{D}_4(\mathbf{C}_4)$ is the two-fold rotation $C_{2z}$.
Therefore,
\begin{eqnarray}
\label{C2z}
  C_{2z}f_{A,B}(\bm{k})=f_{A,B}(-k_x,-k_y,k_z)\nonumber \\
  =-f_{A,B}(k_x,k_y,-k_z)=f_{A,B}(\bm{k}),
\end{eqnarray}
so that $f_{A,B}(k_x,k_y,0)=0$.
Similarly, under a four-fold rotation around the $z$ axis:
$$
C_{4z}f_B(\bm{k})=f_B(k_y,-k_x,k_z)=-f_B(\bm{k}),
$$
therefore $f_B(0,0,k_z)=0$. Also,
$$
C_{4z}f_{^1E,^2E}(\bm{k})=f_{^1E,^2E}(k_y,-k_x,k_z)=\pm
if_{^1E,^2E}(\bm{k}),
$$
hence $f_{^1E,^2E}(0,0,k_z)=0$.

It also follows from Eq. (\ref{C2z}) that $f_{A}(\bm{k})$ and
$f_{B}(\bm{k})$ go to zero at the surface of the Brillouin zone, i.e.
at $k_z=\pm\pi/a$ ($a$ is the lattice constant), because
$(k_x,k_y,\pi/a)$ and $(k_x,k_y,-\pi/a)$ are equivalent points.
In order to take into account the crystal periodicity leading to
the presence of these additional gap zeros, one has to represent the basis
functions as the lattice Fourier series
$f(\bm{k})=\sum_nf_ne^{i\bm{k}\cdot\bm{R}_n}$,
where summation goes over the sites $\bm{R}_n$ of the Bravais lattice of the
crystal. The expansion appropriate for an odd order parameter
has the form
\begin{equation}
\label{lattice_series}
     f(\bm{k})=\sum\limits_n c_n\sin\bm{k}\cdot\bm{R}_n,
\end{equation}
where $\bm{R}_n$ are the sites of a fcc cubic lattice, which cannot
be transformed one into another by inversion.
In the nearest-neighbor approximation, we choose the following
set of $\bm{R}_n$'s: $\{\bm{R}_n\}=(a/2)\{(101),(\bar 101),(011),
  (0\bar 11),(110),(\bar 110)\}$.
Using Table \ref{table2}, we obtain
the basis functions which have symmetry-imposed zeros at the surface
of the Brillouin zone:
\begin{eqnarray*}
     &&f_A(\bm{k})=\sin\frac{k_za}{2}
         \left(\cos\frac{k_xa}{2}+\cos\frac{k_ya}{2}\right)\\
     &&f_B(\bm{k})=\sin\frac{k_za}{2}
         \left(\cos\frac{k_xa}{2}-\cos\frac{k_ya}{2}\right)\\
     &&f_{^1E}(\bm{k})=\cos\frac{k_za}{2}
         \left(\sin\frac{k_ya}{2}+i\sin\frac{k_xa}{2}\right)\\
     &&+\lambda_1\left[e^{\frac{i\pi}{4}}\sin\left(\frac{k_xa}{2}+\frac{k_ya}{2}
     \right)
     -e^{-\frac{i\pi}{4}}\sin\left(\frac{k_xa}{2}-\frac{k_ya}{2}\right)\right]\\
     &&f_{^2E}(\bm{k})=\cos\frac{k_za}{2}
         \left(\sin\frac{k_ya}{2}-i\sin\frac{k_xa}{2}\right)\\
     &&+\lambda_2\left[e^{-
\frac{i\pi}{4}}\sin\left(\frac{k_xa}{2}+\frac{k_ya}{2}\right)
     -e^{\frac{i\pi}{4}}\sin\left(\frac{k_xa}{2}-\frac{k_ya}{2}\right)\right].
\end{eqnarray*}
Here $\lambda_{1,2}$ are arbitrary real constants. The polynomial expressions
for the basis functions from Table \ref{table2} are recovered in the limit of
a ``small'' Fermi surface $\bm{k}\to 0$ [note that $f_B(\bm{k})$ from
Table \ref{table2} can be obtained by including the next-nearest-neighbors
in the expansion (\ref{lattice_series})].  It should be noted that
these nearest-neighbor results give also gap zeros not required by symmetry,
e.g. $f_B(\bm{k})=0$ on the plane $k_x=k_y$. These ``accidental'' zeros
will be removed if higher-neighbor terms are included, but if the
nearest-neighbor terms turn out to be dominant, experiment could find
indications of these accidental zeros.

\subsection*{Case C:\quad $\mathbf{G}_{orb}=\mathbf{D}_3(\mathbf{C}_3) 
             \times\mathbf{I}$\quad ($\bm{M}\parallel[111]$)}
\label{sec:CaseC}

\begin{table}
\caption{\label{table3} The character table and the examples
of the odd basis functions for the irreducible
co-representations of the magnetic point group
$\mathbf{D}_3(\mathbf{C}_3)$.
The overall phases of the basis functions are chosen so that
$K_0C_{2b}f_\Gamma(\bm{k})=f_\Gamma(\bm{k})$.}
\begin{ruledtabular}
\begin{tabular}{|c||c|c|c|c|}
   $\Gamma$  & $E$ & $C_{3\epsilon}$ & $C^{-1}_{3\epsilon}$ &
    $f_\Gamma(\bm{k})$ \\ \hline
   $A$      & 1   & 1 & 1  &  $k_x+k_y+k_z$  \\ \hline
   $^1E$  & 1   & $\omega$ & $\omega^*$ &
                 $e^{-i\pi/3}k_x-k_y+e^{i\pi/3}k_z$ \\ \hline
   $^2E$  & 1   & $\omega^*$ & $\omega$  &
                 $e^{i\pi/3}k_x-k_y+e^{-i\pi/3}k_z$
\end{tabular}
\end{ruledtabular}
\end{table}

The order parameter is given by Eq. (\ref{d1D}), and
the irreducible co-representations are listed in
Table \ref{table3}. The order parameters $\bm{d}_{^1E}(\bm{k})$ and
$\bm{d}_{^2E}(\bm{k})$ vanish
at the points where the line $k_x=k_y=k_z$ crosses the Fermi surface,
but $\bm{d}_A(\bm{k})$ does not have zeros.
The zeros of $\bm{d}_{^1E,^2E}(\bm{k})$ are imposed by symmetry, because
under a three-fold rotation about the axis $\hat{\bm{\epsilon}}$,
$$
  C_{3\epsilon}f_{^1E,^2E}(\bm{k})=f_{^1E,^2E}(k_z,k_x,k_y)=
  e^{\pm 2\pi i/3}f_{^1E,^2E}(\bm{k}),
$$
so that $f_{^1E,^2E}(k_x=k_y=k_z)=0$.

We also give the expressions for the basis functions
of the magnetic point group $\mathbf{D}_3(\mathbf{C}_3)$ in terms of the lattice
Fourier series in the nearest-neighbor approximation:
\begin{eqnarray*}
     f_A(\bm{k})&=&S^{+}_1+S^{+}_2+S^{+}_3
       +i\lambda_1\left(S^{-}_1+S^{-}_2+S^{-}_3\right)\\
     f_{^1E}(\bm{k})&=&\omega^*S^{+}_1+\omega S^{+}_2+S^{+}_3\\
       &&+i\lambda_2\left(\omega^*S^{-}_1+\omega S^{-}_2+S^{-}_3\right)\\
     f_{^2E}(\bm{k})&=&\omega S^{+}_1+\omega^*S^{+}_2+S^{+}_3\\
       &&+i\lambda_3\left(\omega S^{-}_1+\omega^*S^{-}_2+S^{-}_3\right),
\end{eqnarray*}
where $S^{\pm}_1=\sin(k_xa/2\pm k_ya/2)$, $S^{\pm}_2=\sin(k_ya/2\pm k_za/2)$,
$S^{\pm}_3=\sin(k_za/2\pm k_xa/2)$, and $\lambda_{1,2,3}$ are arbitrary
real constants.

\subsection*{Case D:\quad $\mathbf{G}_{orb}=\mathbf{D}_2(\mathbf{C}_2) 
             \times\mathbf{I}$\quad ($\bm{M}\parallel[110]$)}
\label{sec:CaseD}

\begin{table}
\caption{\label{table4} The character table and the examples
of the odd basis functions for the irreducible
co-representations of the magnetic point group
$\mathbf{D}_2(\mathbf{C}_2)$.
The overall phases of the basis functions are chosen so that
$K_0C_{2b}f_\Gamma(\bm{k})=f_\Gamma(\bm{k})$.
$\lambda$ is an arbitrary real constant.}
\begin{ruledtabular}
\begin{tabular}{|c||c|c|c|}
   $\Gamma$  & $E$ & $C_{2a}$ & $f_\Gamma(\bm{k})$ \\ \hline
   $A$      & 1   & 1 &  $k_x+k_y$  \\ \hline
   $B$  & 1   & $-1$ &  $k_z+i\lambda(k_x-k_y)$
\end{tabular}
\end{ruledtabular}
\end{table}

The order parameter is given by Eq. (\ref{d1D}), and
the irreducible co-representations are listed in
Table \ref{table4}.
The order parameter $\bm{d}_{B}(\bm{k})$ does not have zeros, but
$\bm{d}_{A}(\bm{k})$ has the symmetry-imposed lines of zeros where the
plane $k_x=-k_y$ crosses the Fermi surface, because
under a two-fold rotation about the axis $\hat{\bm{a}}$,
\begin{eqnarray*}
  &&C_{2a}f_A(\bm{k})=f_A(k_y,k_x,-k_z)=-f_A(-k_y,-k_x,k_z)\\
  &&=f_A(\bm{k}),
\end{eqnarray*}
so that $f_A(k_x=-k_y)=0$.

The basis functions
of the magnetic point group $\mathbf{D}_3(\mathbf{C}_3)$ in terms of the lattice
Fourier series in the nearest-neighbor approximation:
\begin{eqnarray*} 
     f_A(\bm{k})&=&\cos\frac{k_za}{2}\left(\sin\frac{k_xa}{2}+
     \sin\frac{k_ya}{2}\right)\\
    &&+\lambda_1\sin\left(\frac{k_xa}{2}+\frac{k_ya}{2}\right)\\
     f_B(\bm{k})&=&\sin\frac{k_za}{2}\left(\cos\frac{k_xa}{2}+\cos\frac{k_ya}{2}
     \right)\\
    &&+i\lambda_2\sin\left(\frac{k_xa}{2}-\frac{k_ya}{2}\right),
\end{eqnarray*}
where $\lambda_{1,2}$ are arbitrary real constants.

\subsection*{Case E:\quad $\mathbf{G}_{orb}=\mathbf{C}_1 \times \mathbf{I}$}
\label{sec:CaseE}

The group $\mathbf{C}_1$ has single one-dimensional odd representation,
which is realized by any odd function of $\bm{k}$. Therefore, there are no
symmetry-imposed gap nodes in this case.

\section{Superconducting order parameter at weak spin-orbit coupling}
\label{sec:SOEvolution}

Now let us turn on a weak spin-orbit coupling neglected in the previous
discussion. We shall see that the effect of spin-orbit coupling is two-fold.
First, it mixes together the order parameters on different sheets.
Second, similar to the electromagnetic interaction studied in the
previous sections, it reduces the symmetry of the order parameter and
changes the gap structure on each sheet of the Fermi surface.

In the presence of spin-orbit coupling, the normal state Hamiltonian 
(\ref{Hamilt_0})
contains an extra term:
\begin{equation}
\label{H0so}
 H_{0,s\textrm{-}o}=H_0+\frac{\hbar}{4m^2c^2}\left[\bm{\nabla}U(\bm{r})\times
 \left(\bm{p}+\frac{e}{c}\bm{A}\right)\right]\cdot\bm{\sigma}.
\end{equation}
Spin is no longer a good quantum number and should be replaced by pseudospin
\cite{ueda85}. In contrast to Eq. (\ref{Gfull}), the symmetry group of Eq. 
(\ref{H0so})
cannot be represented as a product of independent orbital and spin groups.
Instead, we have, neglecting the translations,
\begin{equation}
\label{Gfull_so}
    {\cal G}=\mathbf{G}_{s\textrm{-}o}\times U(1),
\end{equation}
where $\mathbf{G}_{s\textrm{-}o}$ consist of rotations which affect both
the orbital and the pseudospin degrees of freedom:
\begin{equation}
\label{s-o_rots}
    R\psi_\alpha(\bm{r})R^{-1}=
    [D^{(1/2)}(R)]_{\alpha\beta}\psi_{\beta}(R^{-1}\bm{r}),
\end{equation}
and also the combined operations $KR$, where $K=C^s_{2e_2}K_0$, so that
\begin{equation}
\label{Kdef}
    K[c\psi_\alpha(\bm{r})]K^{-1}=
    c^*(i\sigma_2)_{\alpha\beta}\psi_{\beta}(\bm{r}),
\end{equation}
where $c$ is an arbitrary $c$-number (note that $K^2=-1$).
The transformation rules for the order parameter become [cf. Eq. 
(\ref{d_rules})]
\begin{equation}
\label{d_rules_so}
\left.\begin{array}{l}
    R d_\alpha(\bm{k})=[D^{(1)}(R)]_{\alpha\beta}d_\beta(R^{-1}\bm{k})\\
    K d_\alpha(\bm{k})=-d^*_\alpha(-\bm{k})=d^*_\alpha(\bm{k}).
\end{array}\right.
\end{equation}
In this case, as shown in Ref. \cite{sam02},
the symmetry of the system is reduced to a magnetic point group
$\mathbf{G}(\mathbf{H})$, and the superconducting order parameter
transforms according to one of the one-dimensional irreducible co-
representations.
Depending on the direction of the magnetization,
$\mathbf{G}(\mathbf{H})=\mathbf{D}_4(\mathbf{C}_4)$, 
$\mathbf{D}_3(\mathbf{C}_3)$,
$\mathbf{D}_2(\mathbf{C}_2)$, or $\mathbf{C}_1$ (in Ref. \cite{sam02},
only the first two cases were studied).
The only difference from the previous Section is that the elements of the
magnetic groups now act simultaneously on the orbital and the spin coordinates, 
see
Eqs. (\ref{s-o_rots}) and (\ref{Kdef}), and one should replace $K_0R$ with $KR$
in the definitions (\ref{GorbCaseB}), (\ref{GorbCaseC}), and (\ref{GorbCaseD}).

Because of the possibility of the interband pairing interactions of the form
$c^\dagger_{\bm{k}\uparrow}c^\dagger_{-\bm{k},\uparrow}
c_{\bm{k}'\downarrow}c_{-\bm{k}',\downarrow}$, the superconductivity is present
on both sheets of the Fermi surface. Instead of Eq. (\ref{dreduced}),
we have the following general expression for the order parameter:
\begin{eqnarray}
\label{d_so}
  &&\bm{d}(\bm{k}) =  \hat{\bm{e}}_+d_-(\bm{k})+\hat{\bm{e}}_-d_+(\bm{k})
        + \hat{\bm{e}}_3 d_3(\bm{k}) \nonumber\\
  &&\approx i\hat{\bm{e}}_+ \sum\limits_{i=1}^{n_{\Gamma_-}}\eta_{-,i}
   f_{\Gamma_-,i}(\bm{k})
   +i\hat{\bm{e}}_- \sum\limits_{i=1}^{n_{\Gamma_+}}\eta_{+,i}
   f_{\Gamma_+,i}(\bm{k}).
\end{eqnarray}
Here $\Gamma_-$ ($\Gamma_+$) label the irreducible
co-representations of $\mathbf{G}(\mathbf{H})$ describing the
orbital symmetry of the order parameter at the pseudospin-up
(pseudospin-down) sheets of the Fermi surface. The choice of these
representations is not arbitrary, because $\hat{\bm{e}}_+d_-$ and
$\hat{\bm{e}}_-d_+$ should have the same symmetry properties. Thus, the
order parameter has $n_{\Gamma_+}+n_{\Gamma_-}$ components:
$(\bm{\eta}_+,\bm{\eta}_-)$, where
$\bm{\eta}_+=(\eta_{+,1},...,\eta_{+,n_{\Gamma_+}})$ and
$\bm{\eta}_-=(\eta_{-,1},...,\eta_{-,n_{\Gamma_-}})$. For the
magnetic groups of interest to us, all co-representations are
one-dimensional, so that $n_{\Gamma_+}=n_{\Gamma_-}=1$. As
discussed in Sec. \ref{sec:SymmOP}, the contribution proportional
to $\hat{\bm{e}}_3$ is small because of the large exchange band
splitting, and is neglected in the second line of 
Eq. (\ref{d_so}).

It is instructive to study the evolution of the order parameter symmetry
in the presence of spin-orbit coupling using the Ginzburg-Landau theory.
Let us start by looking at the first of the effects mentioned in the
beginning of this Section (i.e. the order parameter mixing), using as an example
the vector representation $F_1$ of $\mathbf{O}$ and assuming 
$\bm{M}\parallel[001]$.
For the moment, we neglect the electromagnetic interaction and omit
the gradient terms in the free energy.
At zero spin-orbit coupling, the orbital symmetry is cubic, and
$\Gamma_+=\Gamma_-=F_1$.
It is convenient to use the following set of the basis functions of $F_1$:
\begin{equation}
\label{F1basis}
   f_1(\bm{k})=\frac{k_y+ik_x}{\sqrt{2}},\
   f_2(\bm{k})=\frac{k_y-ik_x}{\sqrt{2}},\
   f_3(\bm{k})=k_z,
\end{equation}
then $\bm{\eta}_\pm=(\eta_{\pm,1},\eta_{\pm,2},\eta_{\pm,3})$,
and the quadratic part of the free energy is
\begin{equation}
\label{GLzero_so}
    F_0=a_+(T-T_{c,+})|\bm{\eta}_+|^2+a_-(T-T_{c,-})|\bm{\eta}_-|^2.
\end{equation}
The critical temperatures $T_{c,-}$ and $T_{c,+}$ for
the spin-up and the spin-down order parameters are different, in general
(we assume that $T_{c,-}>T_{c,+}$).
There are no mixed terms of the form $\eta^*_{+,i}\eta_{-,j}+c.c.$
in Eq. (\ref{GLzero_so}),
because of the spin rotation symmetry $U(1)$. Indeed, under a spin rotation by 
an angle
$\theta$ about $\hat{\bm{e}}_3$, we have $d_\pm\to e^{\pm i\theta}d_\pm$, which can be
interpreted as an operation acting on the order parameter components:
$\bm{\eta}_\pm\to e^{\pm i\theta}\bm{\eta}_\pm$. The mixed terms are
not invariant under such transformations and therefore are not allowed.
This is, of course, the same continuous
symmetry which is responsible for the spin conservation.

Now, if a weak spin-orbit coupling is turned on, we can treat it as a 
symmetry-breaking
perturbation in the phenomenological Ginzburg-Landau functional.
The spin rotations are no longer symmetry elements on their own, and,
in addition to the terms on the right-hand side of Eq. (\ref{GLzero_so}),
the free energy should contain other invariants built from
the components of $(\bm{\eta}_+,\bm{\eta}_-)$.
The magnetic group $\mathbf{D}_4(\mathbf{C}_4)$
is generated by the rotations $C_{4z}$ and the combined operations $KC_{2x}$.
According to (\ref{d_rules_so}), $C_{4z}d_\pm(\bm{k})=\pm id_\pm(C_{4z}^{-
1}\bm{k})$,
$KC_{2x}d_\pm(\bm{k})=d^*_\pm(C_{2x}^{-1}\bm{k})$. In terms of 
$\bm{\eta}_{\pm}$,
we have
\begin{equation}
\label{rules_eta}
\left.\begin{array}{l}
    C_{4z}\eta_{\pm,1}=\mp \eta_{\pm,1}\\
    C_{4z}\eta_{\pm,2}=\pm \eta_{\pm,2}\\
    C_{4z}\eta_{\pm,3}=\pm i\eta_{\pm,3}\\
    KC_{2x}\bm{\eta}_{\pm}=\bm{\eta}^*_{\pm}.
\end{array}\right.
\end{equation}
(Note that, because of our choice of the basis functions and the presence 
of the overall factors $i$ on the right-hand side of Eq. (\ref{d_rules_so}), 
the action of $KC_{2x}$ on the order parameter components is equivalent 
to complex conjugation.) Using Eqs. (\ref{rules_eta}), we obtain quadratic terms
which are invariant under all transformations from $\mathbf{D}_4(\mathbf{C}_4)$ 
and should therefore be added to the free energy (\ref{GLzero_so}):
\begin{eqnarray}
\label{add_terms}
    F_{s\textrm{-}o}=F_0 &+& \sum\limits_{i=1}^3
    \left(\lambda_{+,i}|\eta_{+,i}|^2+\lambda_{-,i}|\eta_{-
,i}|^2\right)\nonumber\\
    &+& \gamma_1(\eta^*_{-,1}\eta_{+,2}+\eta^*_{+,2}\eta_{-,1})\nonumber\\
    &+& \gamma_2(\eta^*_{-,2}\eta_{+,1}+\eta^*_{+,1}\eta_{-,2}).
\end{eqnarray}
The coefficients $\lambda_{\pm,i}$ and $\gamma_{1,2}$ are small at weak spin-
orbit coupling.
The model of Eqs. (\ref{GLzero_so}) and (\ref{add_terms}) can have a rich phase
structure, depending on the relation between the ``bare'' critical temperatures
$T_{c,-}$ and $T_{c,+}$ and other parameters.
In order to work out the whole phase diagram and the structure
of successive superconducting phases, one should include fourth-order terms in
the free energy (\ref{GLzero_so}) and (\ref{add_terms}), which we shall not do 
here. Instead, we concentrate on finding the maximum critical
temperature.

The components $(\eta_{+,1},\eta_{-,2})$, $(\eta_{+,2},\eta_{-,1})$,
$\eta_{+,3}$, and $\eta_{-,3}$ can be considered separately.
For example, the critical temperature for $(\eta_{+,2},\eta_{-,1})$ is given by
\begin{equation}
    T_c=\frac{T_{+,2}+T_{-,1}}{2}+
    \frac{1}{2}\sqrt{(T_{+,2}-T_{-,1})^2+\frac{4\gamma_1^2}{a_+a_-}},
\end{equation}
where $T_{\pm,i}=T_{c,\pm}-\lambda_{\pm,i}/a_\pm$. Both components $\eta_{+,2}$
and $\eta_{-,1}$ are non-zero below $T_c$, so that superconductivity
appears simultaneously on both sheets of the Fermi surface.
The order parameter can be obtained from Eq. (\ref{d_so}):
\begin{equation}
\label{dmixed}
  \bm{d}(\bm{k})=i\hat{\bm{e}}_+\frac{k_y+ik_x}{\sqrt{2}}\eta_{-,1}
   +i\hat{\bm{e}}_-\frac{k_y-ik_x}{\sqrt{2}}\eta_{+,2}.
\end{equation}
At weak spin-orbit coupling and $T_{c,-}>T_{c,+}$, $\eta_{+,2}$
is much smaller than $\eta_{-,1}$: $\eta_{+,2}/\eta_{-,1}\propto\gamma_1$.
The order parameter (\ref{dmixed}) has point nodes at the poles of the Fermi 
surfaces and, according to the classification of Ref.~\cite{sam02},
corresponds to the irreducible co-representation $A$ of
$\mathbf{D}_4(\mathbf{C}_4)$. Similarly, one can derive $T_c$ for the
order parameter $(\eta_{+,1},\eta_{-,2})$ and check that it corresponds to the
co-representation $B$.

The critical temperatures
for $\eta_{\pm,3}$ are $T_{\pm,3}=T_{c,\pm}-\lambda_{\pm,3}/a_\pm$.
The corresponding order parameter $\bm{d}$ still vanishes on one of the
sheets of the Fermi surface, which is an artifact of our model,
based on the representation $F_1$ of $\mathbf{O}$. If one includes
{\em all} representations of the cubic group in the free energy 
(\ref{GLzero_so}),
then the spin-orbit coupling would lead to the appearance of a variety
of quadratic terms which mix together different representations on 
different sheets, similar to Eq. (\ref{add_terms}). In this case,
the order parameter will always be present on both sheets of the Fermi 
surface, and the results of Ref. \cite{sam02} will be recovered.

Now we study how the nodal structure of the superconducting order
parameter on a single sheet (say, the pseudospin-up sheet) evolves
with spin-orbit coupling. We consider only the case
$\bm{M}\parallel[001]$, neglect the electromagnetic interaction,
and start from the representations $A_1$ and $F_1$ of the group
$\mathbf{O}$ at zero spin-orbit coupling. The order parameters
corresponding to $A_1$ is
$\bm{d}_{A_1}(\bm{k},\bm{r})=i\hat{\bm{e}}_+\xi(\bm{r})f_{A_1}(\bm{k})$ 
(see Eq. (\ref{d1D})). The order parameter corresponding to $F_1$ has the
form (\ref{dreduced}) with $\Gamma=F_1$, $n_\Gamma=3$, and the
basis functions given by Eqs. (\ref{F1basis}). The quadratic part
of the Ginzburg-Landau functional is
\begin{equation}
\label{GLA1F1}
    F_0=a_{A_1}(T-T_{A_1})|\xi|^2+a_{F_1}(T-T_{F_1})|\bm{\eta}|^2.
\end{equation}
There are no mixed terms in Eq. (\ref{GLA1F1}) because of the different 
transformation properties of $\xi$ and $\bm{\eta}$ with respect to
the elements of the cubic group. We assume $T_{A_1}>T_{F_1}$, 
so that only $\xi$ is nonzero immediately below the critical temperature.
From Table \ref{table1}, the order parameter $\bm{d}_{A_1}$ has six line
nodes where the planes $k_x=0$, $k_y=0$, $k_z=0$, $k_x=k_y$, $k_y=k_z$, and
$k_z=k_x$ cross the Fermi surface. However, according to Table \ref{table2},
all these gap nodes, except from that
on the plane $k_z=0$, are incompatible with the magnetic symmetry
$\mathbf{D}_4(\mathbf{C}_4)$. Let us now see how the extra nodes disappear
when the spin-orbit coupling is taken into account.

The spin-orbit coupling reduces the cubic symmetry to 
$\mathbf{D}_4(\mathbf{C}_4)$,
whose action on the components
$\bm{\eta}(=\bm{\eta_-})$ is given by Eqs. (\ref{rules_eta}), and on $\xi$ by
\begin{equation}
\label{rules_xi}
\left.\begin{array}{l}
    C_{4z}\xi=-i\xi\\
    KC_{2x}\xi=\xi^*
\end{array}\right.
\end{equation}
[here we used Eq. (\ref{d_rules_so}) and the identities
$f_{A_1}(C^{-1}_{4z}\bm{k})=f_{A_1}(\bm{k})$ and
$f^*_{A_1}(-C^{-1}_{2x}\bm{k})=-f_{A_1}(\bm{k})$].
Since the components $\xi$ and $\eta_3$ have the same transformation
properties under all operations from $\mathbf{D}_4(\mathbf{C}_4)$,
the free energy, which is invariant with respect to the magnetic group,
should contain mixed terms in addition to (\ref{GLA1F1}):
\begin{equation}
   F_{s\textrm{-}o}=F_0+\gamma(\xi^*\eta_3+\eta_3^*\xi),
\end{equation}
where $\gamma$ is small at weak spin-orbit coupling.
The critical temperature is changed compared to $T_{A_1}$:
\begin{equation}
    T_c=\frac{T_{A_1}+T_{F_1}}{2}+
    \frac{1}{2}\sqrt{(T_{A_1}-T_{F_1})^2+\frac{4\gamma^2}{a_{A_1}a_{F_1}}}
\end{equation}
and the order parameter on the pseudospin-up sheet now has the form
\begin{equation}
    \bm{d}(\bm{k})=i\hat{\bm{e}}_+[\xi f_{A_1}(\bm{k})+\eta_3 
f_{F_1,3}(\bm{k})]\propto\hat{\bm{e}}_+k_z.
\end{equation}
This order parameter corresponds to the the co-representation $A$ of
$\mathbf{D}_4(\mathbf{C}_4)$.
Thus, the only line node that survives the presence of the spin-orbit coupling 
is located on the plane $k_z=0$. However, if the spin-orbit coupling is weak, 
then the subdominant component $\eta_3$ is small, and the other five line 
nodes of $f_{A_1}(\bm{k})$ are just slightly filled, so that we shall have 
deep minima in the gap. At not very low temperatures, these ``quasi-nodes'' 
cannot be distinguished experimentally from true line nodes.

\section{Ginzburg-Landau theory for ferromagnetic superconductors}
\label{sec:GL}

We have seen in the previous sections that both the electromagnetic interaction
and the spin-orbit coupling break the cubic symmetry, 
lift the degeneracy of the order parameter, and change the gap structure. 
In addition, the spin-orbit coupling
induces non-zero order parameters on both sheets of the Fermi surface.
The symmetry is reduced to a magnetic
group $\mathbf{D}_n(\mathbf{C}_n)$ ($n=2,3,4$), or $\mathbf{C}_1$. All
co-representations of these groups are one-dimensional, so that the
general form of the order parameter is given by
\begin{equation}
\label{dreduced_so}
  \bm{d}(\bm{k},\bm{r})=i\hat{\bm{e}}_+ f_{\Gamma_-}(\bm{k})\eta_{-}(\bm{r})
   +i\hat{\bm{e}}_- f_{\Gamma_+}(\bm{k})\eta_{+}(\bm{r}).
\end{equation}
The order parameter symmetry should be the same on both sheets, which
means that (i) both components $\eta_{-}$ and $\eta_{+}$ have the same
transformation properties
under the action of the magnetic group elements, and (ii)
there are some restrictions as to the choice of $\Gamma_+$ and $\Gamma_-$,
stemming from the different transformation properties of the
spin vectors $\hat{\bm{e}}_+$ and $\hat{\bm{e}}_-$.
In Table \ref{Table_pairs},
the pairs of orbital co-representations giving rise to the same symmetry
of $\bm{d}$ are listed for all three relevant magnetic groups.
For instance, the order parameter (\ref{dmixed}) corresponds to
$(\Gamma_+,\Gamma_-)=({}^2E,{}^1E)$. The examples of the basis functions
$f_{\Gamma_\pm}(\bm{k})$, which have only the zeros imposed by symmetry,
can be found in Tables \ref{table2}, \ref{table3}, and \ref{table4}.
It is easy to see that the order parameter always has nodes, at least on
one of the sheets of the Fermi surface.

\begin{table}
\caption{\label{Table_pairs} The pairs of orbital co-representations
corresponding
to the same symmetry of the order parameter (\ref{dreduced_so}) on both sheets
of the Fermi surface.}
\begin{ruledtabular}
\begin{tabular}{|c||c|}
   $\mathbf{G}(\mathbf{H})$  &  $(\Gamma_+,\Gamma_-)$ \\ \hline
   $\mathbf{D}_4(\mathbf{C}_4)$  &  $(A,B)$,\ \   $(B,A)$,\ \   
$({}^1E,{}^2E)$,\ \
                          $({}^2E,{}^1E)$  \\ \hline
   $\mathbf{D}_3(\mathbf{C}_3)$  &  $(A,{}^2E)$,\ \   $({}^1E,A)$,\ \
                         $({}^2E,{}^1E)$ \\ \hline
   $\mathbf{D}_2(\mathbf{C}_2)$  &  $(A,B)$,\ \   $(B,A)$ \\
\end{tabular}
\end{ruledtabular}
\end{table}

The Ginzburg-Landau functional contains all possible uniform and gradient terms
which are (i) invariant with respect to $\mathbf{G}(\mathbf{H})$, and
(ii) gauge invariant.
The uniform terms have the same form for all three magnetic groups:
\begin{eqnarray}
\label{FGLuniform}
   F_{uniform} &=& a_+(T-T_+)|\eta_+|^2+a_-(T-T_-)|\eta_-|^2\nonumber\\
   && +\gamma(\eta_+^*\eta_-+\eta_-^*\eta_+)+F_4,
\end{eqnarray}
where $F_4$ is given by
\begin{eqnarray}
    F_4 &=& \beta_1|\eta_+|^4+\beta_2|\eta_-|^4\nonumber\\
     && +\beta_3|\eta_+|^2|\eta_-|^2+\beta_4(\eta_+^{*,2}\eta_-^2+
                                     \eta_-^{*,2}\eta_+^2)\\
     && +(\beta_5|\eta_+|^2+\beta_6|\eta_-|^2)(\eta_+^*\eta_-+\eta_-^*\eta_+).
     \nonumber
\end{eqnarray}
The coefficients $\gamma,\beta_4,\beta_5,\beta_6$ vanish at zero spin-orbit 
coupling, due to the spin rotation symmetry. 

The gradient terms are different for different magnetic groups.
For $\mathbf{G}(\mathbf{H})=\mathbf{D}_4(\mathbf{C}_4)$,
\begin{eqnarray}
\label{GLD4C4}
   F_{grad} &=& K^+_1|\bm{D}_\perp\eta_+|^2+K^+_3|D_z\eta_+|^2 \nonumber\\
   && +K^-_1|\bm{D}_\perp\eta_-|^2+K^-_3|D_z\eta_-|^2\nonumber\\
   && +K_4[(\bm{D}_\perp\eta_+)^*(\bm{D}_\perp\eta_-)+c.c.]\nonumber\\
   && +K_6[(D_z\eta_+)^*(D_z\eta_-)+c.c.],
\end{eqnarray}
where $\bm{D}_\perp=(D_x,D_y)$. The coefficients $K_4$ and $K_6$ vanish 
in the absence of spin-orbit coupling.

In the case of
$\mathbf{G}(\mathbf{H})=\mathbf{D}_3(\mathbf{C}_3)$, it is convenient to
make a change of coordinates after which $\hat{\bm{z}}$ is directed along
[111]: $\bm{r}\to\bm{r}'=R\bm{r}$, where
$R$ is the matrix of a three-dimensional rotation by an angle
$\theta=\arccos(1/\sqrt{3})$
about the axis $\bm{B}$. Omitting the primes, the gradient terms in the new
coordinates have the same form as Eq. (\ref{GLD4C4}).

Finally, for $\mathbf{G}(\mathbf{H})=\mathbf{D}_2(\mathbf{C}_2)$, it is
convenient to rotate the coordinates in such a way that
$\hat{\bm{z}}$ is directed along [110]: $\bm{r}\to\bm{r}'=R\bm{r}$, where
$R$ is the matrix of a three-dimensional rotation by an angle $\theta=\pi/2$
about the axis $\bm{B}$. In this case, the gradient terms have the following
form:
\begin{eqnarray}
\label{GLD2C2}
   &&F_{grad} = K^+_1|D_x\eta_+|^2+K^+_2|D_y\eta_+|^2+K^+_3|D_z\eta_+|^2 
\nonumber\\
   &&\quad+K^-_1|D_x\eta_-|^2+K^-_2|D_y\eta_-|^2+K^-_3|D_z\eta_-|^2\nonumber\\
   &&\quad+[K_4(D_x\eta_+)^*(D_x\eta_-)+K_5(D_y\eta_+)^*(D_y\eta_-)\nonumber\\
   &&\quad+K_6(D_z\eta_+)^*(D_z\eta_-)+c.c.].
\end{eqnarray}
Because of the choice of coordinates, $\bm{M}=M\hat{\bm{z}}$ and
$\bm{B}=B\hat{\bm{z}}$ in all three cases. The coefficients $K_4,K_5,K_6$
vanish in the absence of spin-orbit coupling.

If the ferromagnetic magnetization is not directed along a high symmetry axis,
then $\mathbf{G}(\mathbf{H})=\mathbf{C}_1$. In this case, the only symmetry
element is the unity operation, and  
the gradient terms contain all possible real combinations of the components of 
$\bm{D}$ and $\eta_\pm$. We shall not give these (rather cumbersome) expressions
here. 

The Ginzburg-Landau functionals listed above can be used for deriving the
phase diagram of a cubic ferromagnetic superconductor, which can be quite 
complex. In particular, one cannot exclude the possibility of extra phase 
transitions in the superconducting state.

\section{Conclusions}
\label{sec:Concl}

We have studied the symmetry of the superconducting order parameter 
in a cubic ferromagnetic superconductor. An experimental example is provided 
by ZrZn$_2$.
Because of the anti-unitarity of time reversal, the usual symmetry analysis
of possible superconducting states (see Refs. \cite{vol85}, \cite{sigr91}, and 
\cite{min99}) is not applicable. 
In a metallic ferromagnet, when both the electromagnetic
interaction and the spin-orbit coupling are present, the order parameter
symmetry evolves from that appropriate for the cubic group $\mathbf{O}_h$ 
to one of the magnetic point groups, which is studied here using the 
phenomenological Ginzburg-Landau theory.
It is shown that the order parameter corresponds to one of 
the irreducible co-representations of the magnetic group, 
and has two components, which describe pairing on the exchange-split sheets
of the Fermi surface, see Eq. (\ref{dreduced_so}).
It should be noted that our results follow from general symmetry considerations
and do not depend on the nature of ferromagnetism in the normal state 
(itinerant vs localized moments) or the mechanism of superconducting pairing.

We have determined the $\bm{k}$-dependence of the order parameter imposed by the 
magnetic symmetry for all possible directions of the ferromagnetic magnetization,
see Table \ref{Table_pairs}.
The most remarkable result is that there should always be zeros in the energy gap, 
either point nodes or line nodes or both, at least on one of the sheets of the 
Fermi surface, when $\bm{M}$ is directed along any of the high symmetry axis 
of the cubic lattice. 
These nodes should give rise to a power-law behavior of the thermodynamic
and kinetic characteristics \cite{min99}. It is expected
that such experimental techniques as ultrasonic attenuation measurements
in the superconducting state might be especially useful in determining the 
detailed structure of the order parameter (a discussion of this can be found 
in Ref. \cite{sam02}). It should be noted that, if the electromagnetic
and the spin-orbit interactions are weak, then the gap nodes appropriate for the
underlying cubic symmetry would manifest themselves as deep minima of the gap, 
which would also have to be taken into account when analyzing 
the experimental data.

The situation might be complicated by the presence
of additional phase transitions in the superconducting state, which is
a common feature of the systems with multi-component order parameters.
Because of the complexity of the Ginzburg-Landau functionals derived in 
Sec. \ref{sec:GL}, the number of possible scenarios with different predictions
for experiment is quite large. 
In our view, it is still premature to discuss specific models, 
because of the lack of experimental data in the superconducting phase of 
ZrZn$_2$.

\section*{Acknowledgements}

We are pleased to acknowledge the
support from the Canadian Institute for Advanced Research and from
the Natural Sciences and Engineering Research Council of Canada.

\end{document}